\documentclass{llncs}

\usepackage{url}            	
\usepackage{booktabs}     
\usepackage{amsfonts}     
\usepackage{nicefrac}       
\usepackage{microtype}    
\usepackage{mathtools}
\usepackage{wrapfig}
\usepackage{paralist}

\newcommand{\reals}{\mathbb{R}}
\newcommand{\nnreals}{\mathbb{R}_{\geq 0}}
\newcommand{\U}{\mathcal{U}}
\newcommand{\R}{\mathcal{R}}

\newcommand{\toolname}{CODEV\xspace}

\usepackage{tikz}
\usetikzlibrary{patterns}
\usepackage{subcaption}
\usetikzlibrary{shapes.geometric, arrows}

\usepackage[mathscr]{euscript}
\usepackage{nameref}
\usepackage{algorithm2e}
\usepackage{adjustbox}
\usepackage{mathpartir,xparse}
\usepackage{bbm}

\begin{document}

\title{\toolname: Automated Model Predictive Control Design and Formal Verification (Tool Paper)}
\titlerunning{\toolname}  
%
\author{Nicole Chan \and Sayan Mitra}

\authorrunning{N. Chan and S. Mitra} 

\tocauthor{Nicole Chan and Sayan Mitra}

\institute{
  Coordinated Science Laboratory,\\
  University of Illinois at Urbana-Champaign\\
  \email{\{nschan3,mitras\}@illinois.edu}
}

\maketitle              

\begin{abstract}
We present \toolname{}, a Matlab-based tool for verifying systems employing Model Predictive Control (MPC).
The MPC solution is computed offline and modeled together with the physical system as a hybrid automaton, whose continuous dynamics may be nonlinear with a control solution that remains affine. 
While MPC is a widely used synthesis technique for constrained and optimal control in industry, our tool provides the first automated approach of analyzing these systems for rigorous guarantees of safety.
This is achieved by implementing a simulation-based verification algorithm for nonlinear hybrid models, with extensions tailored to the structure of the MPC solution. Given a physical model and parameters for desired system behavior (i.e. performance and constraints), \toolname{} generates a control law and verifies the resulting system will robustly maintain constraints. We have applied \toolname{}  successfully to a set of benchmark examples, which illuminates its potential to tackle more complex problems for which MPC is used.
\end{abstract}

\section{Introduction}
\label{sec:intro}

Model Predictive Control (MPC) has become the de facto technique for obtaining high performance controllers in constrained systems~\cite{borrelli2017predictive,camacho2012model,rawlings2009model}.
It is extensively used in industry~\cite{QIN_survey,GARCIA_survey}.
%
Given a linear discrete-time model of the plant dynamics, MPC samples the current state and solves an open-loop optimal control problem over a finite time horizon. 
The first step of the resulting control solution is applied, and the control input for the next step is computed by recomputing the optimization problem with a newly-sampled state and shifted horizon.
The strengths of this approach include optimal performance, appropriately constrained behavior, and dynamic error correction.

However, a control solution, if one exists, is only guaranteed to obey constraints for the immediate prediction horizon under the assumption that the plant model is correct.
\toolname{} addresses this problem by formally verifying the safety of the resulting closed-loop system.
This is achieved by solving an \emph{explicit MPC} formulation to obtain a piecewise affine control law~\cite{Bemporad_eMPC,Alessio2009}. 
This requires solving multi-parametric programs, for which we use the Multi-Parametric Toolbox (MPT)~\cite{MPT3}.
For the verification step, the closed-loop system is viewed as a hybrid automaton, where each discrete mode of operation corresponds with a partition of the state-space generated by MPC, and the continuous dynamics in each mode corresponds with the plant dynamics with the controller. The plant dynamics here may be nonlinear, facilitating more accurate representation of the true system behavior than what can be specified in the MPC problem. The tool takes this hybrid model, a compact set of initial states, a finite time horizon, and a set of safety properties and certifies if the system will behave within the constraints of these safety properties under these conditions. 

The underlying simulation-based verification algorithm is identical to the that implemented in C2E2~\cite{Fan_CAV16}---one of few stable tools available for tackling nonlinear hybrid systems. 
However, an explicit MPC solution can easily contain thousands of sub-functions (i.e. thousands of discrete modes in the automaton,) which is impractical to capture in the static model specification input files required by C2E2 and many existing verification tools~\cite{DBLP:conf/cav/FrehseGDCRLRGDM11,chen2013flow,DuggiralaMVP15,gao2013dreal}. 
Thus \toolname{} is implemented in the same environment for which we have support to solve explicit MPC: Matlab. 
The tool couples the control synthesis and verification processes to provide a strengthened control design framework, for complex, real-world applications.

\paragraph{Related work.}
Despite the popularity of MPC, there is surprisingly little research on formal verification of MPC-based control systems. As previously mentioned, there is no tool that can directly perform bounded safety verification of MPC-based control systems. 
Further, we found no case-studies (not even academic ones) that report of applying verification to an MPC-based system.
In~\cite{Bemporad2006HSCC}, an event-driven model predictive control scheme with verified closed-loop convergence have been proposed for a restricted class of integral hybrid automata. 
%
There are several works that use MPC-based ideas for solving the related but different problem of  controller synthesis~\cite{raman2014model,RamanDMMSS17,wongpiromsarn2011tulip}. 


\section{Verified MPC Approach}
\label{sec:approach}

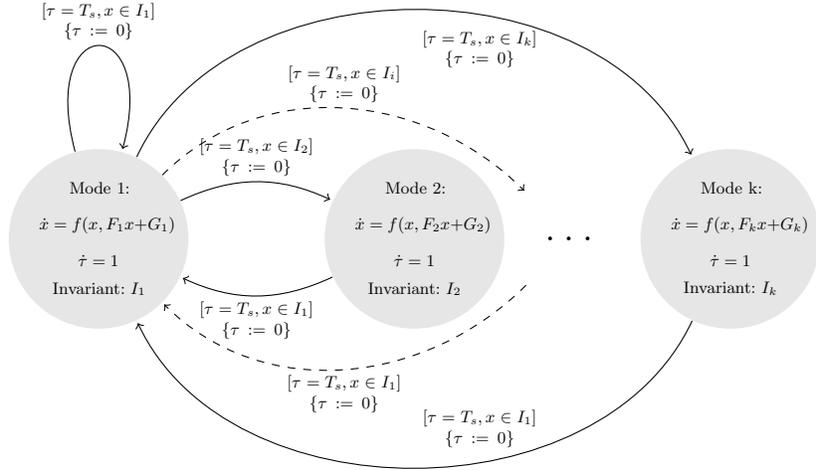
\begin{figure}
	\centering
	\begin{tikzpicture}[scale=0.7]
		\tikzstyle{every node}=[circle,text width=2.25cm,fill=black!10,align=center,scale=0.7]
		\draw (-6,0) node (p1) {Mode 1: $$\dot{x} = f(x,F_1 x+ G_1)$$ $$\dot{\tau}=1$$ Invariant: $I_1$};
		\draw (0,0) node (p2) {Mode 2: $$\dot{x} = f(x,F_2 x+ G_2)$$ $$\dot{\tau}=1$$ Invariant: $I_2$};
		\draw (6,0) node (pk) {Mode k: $$\dot{x} = f(x,F_k x+ G_k)$$ $$\dot{\tau}=1$$ Invariant: $I_k$};
		\node[fill=none,font=\Huge] (dots) at (3,0) {\ldots};
		\tikzstyle{every node}=[fill=none,align=center,text width=2.25cm,scale=0.7]
		\draw [ shorten >=1pt,->] (p1) to [out=25,in=155] node[above] {[$\tau=T_s, x\in I_2$]\\ \{$\tau:=0$\}} (p2);
		\draw [ shorten >=1pt,->] (p2) to [out=-155,in=-25] node[below] {[$\tau=T_s, x\in I_1$]\\ \{$\tau:=0$\}} (p1);
		\draw [ shorten >=1pt,->,dashed] (p1) to [out=45,in=135] node[above] {[$\tau=T_s, x\in I_i$]\\ \{$\tau:=0$\}} (dots);
		\draw [ shorten >=1pt,->,dashed] (dots) to [out=-135,in=-45] node[below] {[$\tau=T_s, x\in I_1$]\\ \{$\tau:=0$\}} (p1);
		\draw [ shorten >=1pt,->] (p1) to [out=65,in=115] node[below=0.75cm,right] {[$\tau=T_s, x\in I_k$]\\ \{$\tau:=0$\}} (pk);
		\draw [ shorten >=1pt,->] (pk) to [out=-115,in=-65] node[above=0.75cm,right] {[$\tau=T_s, x\in I_1$]\\ \{$\tau:=0$\}} (p1);
		\draw [ shorten >=1pt,->] (p1) to[loop above] node[] {[$\tau=T_s, x\in I_1$]\\ \{$\tau:=0$\}} (p1);
	\end{tikzpicture}
	\caption{\scriptsize Hybrid automaton model after computing the piecewise affine control law~\eqref{eq:affinecontrol}, with invariants $I_i \subseteq \reals^n$ corresponding to the partitions given by MPC. A timer variable $\tau$ captures the periodic updates of MPC. Transitions are taken when the guard denoted by $[]$ is satisfied and the reset action in $\{\}$ is applied. Only transitions to/from Mode 1 are shown for brevity.}
	\label{fig:hymod}
\end{figure}

Let the dynamics of the physical system be described by:
\begin{subequations}
\label{eq:closedloop}
\begin{align}
	\dot{x}(t) &= f(x(t),u(t)) \label{eq:plantmodel}\\
	u(t) &= F_i x(t) + G_i, i=1,2,...,k, \label{eq:affinecontrol}
\end{align}
\end{subequations}
with state $x\in\reals^n$ evolving according to $f:\reals^n\to\reals^n$, a Lipschitz continuous function, 
and a piecewise affine control input $u\in\reals^m$ given by the MPC solution. The MPC solution provides $k$-partitions of the state-space, each encoded by the control law parameters $F_i\in\reals^{m\times n}$ and $G_i\in\reals^{m\times1}$.
%
This system lends naturally to a hybrid model, as shown in Figure~\ref{fig:hymod}. 

The controller should drive the state to a reference point $x_{\mathit{ref}}=0$, which we fix to be the origin without loss of generality. It should do so in a way that optimizes a cost function $J: \reals^n\times\reals^{m\times N}\to\nnreals$ and avoids visiting any unsafe sets $\U\subseteq\reals^n$. The verification problem is to rigorously prove that the system cannot reach any state in $\U$ under bounded initial states and time horizon.

\paragraph{MPC design approach.} MPC accounts for the cost function and some or all of the unsafe states directly. We fix a prediction horizon of $N$-discrete steps (sampled with a period of $T_s$). The MPC problem with constraints  $[u_{min}, u_{max}]$ on input and $[x_{min}, x_{max}]$ on state is then the following LP:
\begin{subequations}
\label{eq:mpc}
\begin{align}
	\min_{U=\{u_k,...,u_{k+N-1}\}} \{ J(U,x(k)) &= \sum_{i=0}^{N-1} \left( ||Qx(k+i)||_{\infty} +  ||Ru(k+i)||_{\infty} \right) \}\label{eq:costFun}\\
	\text{s.t. }
	& u_{min} \leq u(k+i) \leq u_{max}, i = 0,1,...,N-1 \label{eq:optConstraint1}\\
	& x_{min} \leq x(k+i) \leq x_{max}, i = 1,2,...,N \label{eq:optConstraint2}\\
	& x(k+i+1) = A_d x(k+i) + B_d u(k+i), i\geq 0 \label{eq:discreteLTI}\\
	& u(k+i) = 0, i\geq N 
\end{align}
\end{subequations}
The constraints include a linear discrete-time approximation of the continuous-time dynamics~\eqref{eq:plantmodel} in~\eqref{eq:discreteLTI}, with $A_d\in\reals^{n\times n},B_d\in\reals^{n\times m}$. The cost~\eqref{eq:costFun} uses the $\infty$-norm here, but the 1-norm can also be used while preserving the LP structure. 
See~\cite{Bemporad_eMPC,Bemporad_mpqp} for a 2-norm (and quadratic programming) approach.
This optimization problem is re-casted into a multi-parametric linear program (mp-LP), whose solution gives the partitions of the state-space and corresponding control inputs (the $F_i$'s and $G_i$'s  in~\eqref{eq:affinecontrol}). 

\paragraph{Verification problem and approach.}
Given the initial state of the system $x(0)$ is  in the compact set $\Theta\subseteq\reals^n$ and a verification time horizon $T_v$, the goal is to formally show that the system in Figure~\ref{fig:hymod} (with possibly nonlinear dynamics~\eqref{eq:closedloop}) cannot reach a state $x(t)\in \U$ for all $0\leq t\leq T_v$. Our  approach is to compute an over-approximation of all reachable states, the \emph{reachtube} $\R_{[0,T_v]}$ using the algorithm of~\cite{Fan_CAV16,Fan_ATVA15} and checking that $\R_{[0,T_v]} \cap \U =\emptyset$.

\section{Usage and Implementation}
\label{sec:implement}

%
%
%
\toolname is implemented in Matlab and uses the third-party MPT3 Toolbox~\cite{MPT3}. The user interacts with the tool by providing an input file that is structured as described in Section~\ref{ssec:inputFile}, and passing its function handle to the main function \texttt{verifyMPC}. This function returns a boolean indicating whether the problem is verified to be \textsc{robustly safe} (\texttt{safeFlag=1}) or otherwise unknown (\texttt{safeFlag=0}). It also returns the corresponding reachtube \texttt{reach}---a temporally-sorted array of polyhedron-objects---that is empty if a safe result cannot be obtained. The polyhedron data type we will refer to is the \texttt{Polyhedron} class provided by MPT3. The output \texttt{reach} can be plotted in Matlab as usual: \texttt{plot(reach)}, if the model is 2-dimensional. Otherwise, the \texttt{projection} method may be used to plot in dimensions $i,j$ as follows: \texttt{plot(reach.projection([i,j]))}. Figure~\ref{fig:architecture} illustrates the high-level structure of the tool implementation.

\tikzstyle{task} = [rectangle, draw, fill=white, text width=7cm, text centered, node distance=1.5cm, minimum height=4em]
\tikzstyle{block} = [rectangle, thick,draw, fill=white, text width=4cm, text centered, rounded corners]
\tikzstyle{line} = [draw, -latex']
\tikzstyle{cloud} = [draw, ellipse,fill=blue!20, node distance=1.5cm,minimum height=2em,text centered]

\begin{figure}
\centering
\scalebox{0.8}{
\begin{tikzpicture}[node distance = 1.5cm,auto]
    \node [block] (init) {Model input script};
    \node [rectangle, thick,draw, fill=black!10,text width=12cm, rounded corners, minimum height=6.5cm,below of=init, node distance = 3.75cm] (tool) {};
    \node [above=2.75cm, left=4.25cm] at (tool) {\toolname};
    \node [task, below of=init,minimum height=1.5cm,node distance=1.5cm] (modelGen) {};
    \node [above=0.25cm] at (modelGen) {Backend model generation};
    \node [cloud, below=-0.1cm] at (modelGen) {Solve for MPC};
    \node [task, below of=modelGen,minimum height=4cm,node distance = 3.25cm, text width=10cm] (core) {};
    \node [above=1.5cm] at (core) {Core verification engine};
    \node [cloud,left of=core, node distance = 3.25cm, text width=1.5cm,below=-1em] (partition) {Partition initial set};
    \node [block,right=1.5cm,below=-1.25cm,text width=5.75cm, minimum height=3cm] at (core) (reach) {};
    \node [above=1cm] at (reach) {Compute Reachtube};
    \node [cloud, left of=reach,node distance=1.5cm,text width=1.5cm] (computePost) {Compute Post};
    \node [cloud, right of=reach,node distance=1.5cm,text width=1.5cm] (checkSafe) {Check safety};
    \node [block,below of=tool, node distance=3.75cm] (result) {Safety result and reachtube};
    \path [line] (init) -- (modelGen);
    \path [line] (modelGen) -- (core);
    \path [line] (core) -- (result);
    \path [line] (partition) -- (computePost);
    \path [line] (computePost) -- (checkSafe);
    \node [below=1cm] at (checkSafe) (test) {};
    \coordinate [below=1cm] (pt1) at (checkSafe);
    \draw (checkSafe) -- node [right=0.45cm,below=0.2cm,font=\itshape] {Robustly Safe} (pt1);
    \path [line] (pt1) -|  (computePost);
    \coordinate [above=1cm] (pt2) at (checkSafe);
    \draw (checkSafe) -- node [right=-0.1cm,font=\itshape] {Otherwise} (pt2);
    \path [line] (pt2) -| (partition);
\end{tikzpicture}
}
\caption{\scriptsize Overview of the tool's structure.}
\label{fig:architecture}
\end{figure}
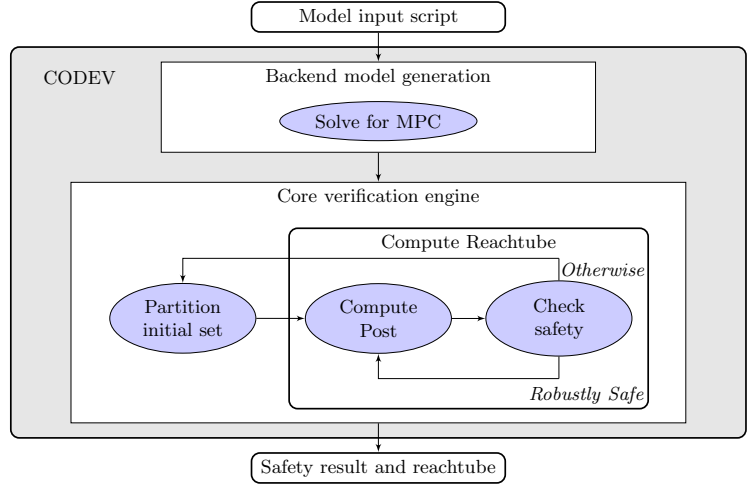

\subsection{Input Specification}
\label{ssec:inputFile}

In order to fully specify the safety verification problem, three main components are needed:
\begin{inparaenum}[(i)]
\item dynamics of the system of interest (Eq.~\eqref{eq:closedloop}), 
\item parameters for controller design/synthesis (Eq.~\eqref{eq:mpc}), and 
\item safety verification parameters ($\U, \Theta, T_v$).
\end{inparaenum}
%
All of these components are stored in a custom struct, which is instantiated in the \emph{model generation} block of Figure~\ref{fig:architecture} according to the user input file.
Let us refer to this object as \texttt{inParams}, which must include the following fields: \texttt{flowEq, modelJac, unsafeStates, initStates, vPar, MPCprob}, and optionally \texttt{MPCsol}. 
The physical dynamics are represented in \texttt{flowEq} and \texttt{modelJac}. The verification parameters are stored in \texttt{unsafeStates}, \texttt{initStates}, and \texttt{vPar}. The parameters for MPC design are specified in \texttt{MPCprob}. However, an explicit MPC solution can be solved separately and stored in \texttt{MPCsol}, in which case \texttt{MPCprob} can be set to an empty object.

\subsection{Model Generation}
\label{ssec:modelGen}
The \emph{model generation} module receives a function handle, calls the function to instantiate the input object \texttt{inParams}, parses the input fields for correctness and calls the MPC solver if no \texttt{MPCsol} field is provided. 
Then the solution is stored in the added \texttt{MPCsol} field. The struct \texttt{inParams.MPCsol} contains the following fields: array of \texttt{Polyhedron} partitions of the state-space \texttt{Pn}, cell-array of matrices \texttt{Fi}, and cell-array of matrices \texttt{Gi}.
For index \texttt{i}, if the state of the system is contained in region \texttt{Pn\{i\}}, then its corresponding control law~\eqref{eq:affinecontrol} has parameters \texttt{Fi\{i\}}, \texttt{Gi\{i\}}.

\subsection{Verification Algorithm}
\label{ssec:verify}

In the \emph{core verification} component, initial set $\Theta$ (or \texttt{inParams.initStates}) is first partitioned into $\theta_i, i=1,2,...,k$, such that $\cup_{i} \theta_i \supseteq \Theta$ and each $\theta_i$ is contained in a single discrete mode of the hybrid model.
For each $\theta_i$, we apply the simulation-based reachability algorithm given in~\cite{Fan_ATVA15}, implemented in a method called \texttt{computeReach}. 
Algorithm 1 from~\cite{Fan_ATVA15} is implemented in a method named \texttt{computePost}, with lines 11-15 replaced with an \texttt{else} block that returns an unsafe result (and empty reachtube).
Here, the unsafe result (\textsc{MaxPart} or \textsc{Infeasible}) only indicates that we cannot show the system is \textsc{RobustlySafe}, whereas in the original algorithm, this result indicates that a counterexample for safety has been found. 
Specifically, \textsc{MaxPart} indicates the program exiting when the maximum number of allowable partitions in the verification algorithm is reached, and \textsc{Infeasible} indicates the program exiting because the over-approximated reachsets intersect with regions outside of the MPC solution (feasible region). If a simulation trace could reach the infeasible region, we would consider a counterexample for safety as well.
When \texttt{computePost} returns a safe result, the reachtube $\R_{[t,t+T_s]}$ under a fixed discrete mode is also returned.
Due to the MPC framework, transitions are restricted to occur only periodically (every $T_s$-time at most), so \texttt{computeReach} calls \texttt{computePost} to compute each subset of the reachtube over $T_s$ intervals of time in order to construct the full reachtube $\R_{[0,T_v]}$.

%

\section{Experimental Results}
\label{sec:results}

\toolname{} has been successfully applied to a modest set of examples: the \emph{double integrator}~\cite{Bemporad_eMPC}, \emph{navigation} system~\cite{C2E2_nav_example}, and \emph{cruise control} and \emph{magnetic pointer}~\cite{DigitalControlSynthesis_examples}.
The results presented in Table~\ref{table:results} summarizes their input parameters, 
\toolname{} results, and runtime.
See their input files included with the tool for more details. 

\begin{table}[h]
\centering
    \begin{tabular}{ | l | c  | c | c | p{1.5cm} | c | p{1.5cm} | c |}
    \hline
      Benchmark & $n$ & $N$ & $k$ & MPC solve time & $T_v$ & verification runtime & result \\ \hline
    \emph{double int} & 2  & 2 & 20 & 1.25s & 9s & 165.10s & RobustSafe\\ \hline
    \emph{navigation} & 4  & 2 & 580 & 50.08s & 2s & 102.28s  & RobustSafe \\ \hline
    \emph{cruise control} & 1 &10 & 3 & 5.31s & 10s & 19.13s & RobustSafe \\ \hline
    \emph{magnetic ptr A} & 3 & 4 & 1646 & 154.68s & 1.1s & 33.39s & RobustSafe \\ \hline
    \emph{magnetic ptr B} & 3  & 4 & 1646 & 154.68s & 2s & 70.96s & MaxPart \\ \hline
    \emph{magnetic ptr C} & 3  & 2 & 212 & 12.45s & 2s & 28.65s & Infeasible \\
    \hline
    \end{tabular}
\caption{\scriptsize Parameters for MPC formulation, its computation time, verification horizon, and its runtime (excluding MPC computation time). In all cases: $m=1$; $Q,R$ are identity matrices of dimensions $n,m$ respectively; and $T_s=1$s.}
\label{table:results}
\end{table}

We use the \emph{magnetic pointer} case study to demonstrate some of the challenges with obtaining robust safety results using \toolname{}. We are able to obtain a safe result by restricting the verification horizon to a short length. 
Otherwise, the increasing number of control updates results in a growing number of possible transitions taken at each update. This results in compounding over-approximations of reachable states. 
We observe two resulting sources of failure:
1. the over-approximated regions may intersect with unsafe states, or 2. these regions may intersect with infeasible regions in the MPC solution.

Thus, care should be taken in formulating the MPC problem~\eqref{eq:mpc} to ensure the solution space is large enough (e.g. by relaxing constraints or increasing prediction horizon). 
For verification, the user may design a set of smaller problems (e.g. tighter verification horizon or initial set) to cover a larger verification problem.


\section{Conclusions}
\label{sec:conclusion}
We presented \toolname{}, the first tool for formally verifying systems employing Model Predictive Control (MPC). It implements a simulation-based verification algorithm for nonlinear hybrid models, with extensions tailored to handle MPC-based control systems. The benchmarks used in this paper are modest in size, but our experimental results establish the promise of this approach for more complex and realistic control systems.

The next steps for further developing \toolname{} include exploring more efficient backend mp-LP solvers, implementing support for the QP formulation~\cite{Bemporad_mpqp}, and making the tool more usable by automatically solving for the Jacobian and discrete-time linear model approximation within the \emph{model generation} component (Fig.~\ref{fig:architecture}) so the user need not provide these in the input file.


\section*{Acknowledgments}
This work is supported by NSF grant CNS 1629949.

\bibliographystyle{abbrv}
\bibliography{Chan_cav,sayan1}
\end{document}